\documentclass{Interspeech}
\usepackage{cite}
\usepackage{amsmath,amssymb,amsfonts}
\usepackage{algorithmic}
\usepackage{url}
\usepackage{graphicx}
\usepackage{textcomp}
\usepackage{booktabs}
\usepackage{xcolor}
\usepackage{color, colortbl}
\newcommand{\ci}[1]{\tiny{±#1}}
\definecolor{Gray}{gray}{0.94}

\definecolor{color1}{HTML}{FFF8EF}
\definecolor{color2}{HTML}{FFD8AD}
\definecolor{color3}{HTML}{F7E9DB}
\usepackage{amsthm}
\usepackage{mathtools}
\usepackage{thmtools}
\usepackage[mathscr]{eucal}

\interspeechcameraready

\title{AxLSTMs: learning self-supervised audio representations with xLSTMs}

\author[affiliation={1,2}]{Sarthak}{Yadav}
\author[affiliation={1,3}]{Sergios}{Theodoridis}
\author[affiliation={1,2}]{Zheng-Hua}{Tan}

\affiliation{Department of Electronic Systems}{Aalborg University}{Denmark}
\affiliation{}{Pioneer Centre for Artificial Intelligence}{Denmark}
\affiliation{Department of Informatics and Telecommunications }{National and Kapodistrian University of Athens}{Greece}
\email{sarthaky@es.aau.dk, sthe@es.aau.dk, zt@es.aau.dk}

\usepackage{comment}

\begin{document}

\maketitle

\begin{abstract}
    
While the transformer has emerged as the eminent neural architecture, several independent lines of research have emerged to address its limitations. Recurrent neural approaches have observed a lot of renewed interest, including the extended long short-term memory (xLSTM) architecture, which reinvigorates the original LSTM. However, while xLSTMs have shown competitive performance compared to the transformer, their viability for learning self-supervised general-purpose audio representations has not been evaluated. This work proposes Audio xLSTM (AxLSTM), an approach for learning audio representations from masked spectrogram patches in a self-supervised setting. Pretrained on the AudioSet dataset, the proposed AxLSTM models outperform comparable self-supervised audio spectrogram transformer (SSAST) baselines by up to 25\% in relative performance across a set of ten diverse downstream tasks while having up to 45\% fewer parameters.
\end{abstract}

\section{Introduction}

In lieu of their excellent generalisation capabilities and domain and data agnostic nature, transformers \cite{vaswani2017attention} and their successors have seen widespread adoption. Furthermore, transformers, together with masked predictive modelling, have emerged as a key driving force behind several prominent advancements in the realm of unsupervised and self-supervised representation learning for NLP \cite{bert2019}, computer vision \cite{he2022masked} and audio and speech processing \cite{baevski2020wav2vec, hsu2021hubert, gong2022ssast, huang2022masked, chen2022wavlm}. 
Recently, however, finding alternatives to scaled dot-product attention has garnered significant interest, with a lot of emphasis on finding sub-quadratic appromixations of the attention operation \cite{katharopoulos2020transformers} as well as token mixing \cite{tolstikhin2021mlp}. 

Before the advent of transformers, sequence modeling was predominantly done using recurrent neural networks (RNNs). 
Recurrent models offer several advantages over transformers: they scale linearly with respect to sequence length and they have lower runtime memory requirements since storing the entire key-value (KV) cache is not necessary. State-space models (SSMs) \cite{gu2022efficiently}, which are a family of sequence models that lie at the intersection of convolutional neural networks, RNNs and classical state spaces, are the most popular amongst recent recurrent approaches. Several variants of SSMs have been proposed, showing competitive performance and scalability versus transformers in several domains, including long sequence modelling \cite{gu2023mamba}, computer vision \cite{zhu2024vision} as well as audio \cite{yadav2024audiomambaselectivestate}. 

However, prior to transformers and state-space models, LSTMs \cite{hochreiter1997long} were the go to neural architecture for sequence modeling. Compared to transformers, LSTMs suffer from several key drawbacks: (i) inability to revise storage decisions, (ii) compressing information into a scalar cell state, which impacts performance on rare input tokens, and (iii) memory mixing and the resulting lack of parallelizability. Recently, \cite{beck2024xlstmextendedlongshortterm} proposed the extended long-short term memory (xLSTM) neural architecture, which revitalises the original LSTM architecture, leveraging the latest techniques and tricks learned from years of transformer and large language modelling research. xLSTM incorporates exponential gating,  improved normalization and stabilization techniques while removing traditional memory mixing, resulting in  two new fundamental building blocks: sLSTM and mLSTM. xLSTMs have demonstrated on-par or better performance than transformers for large language models \cite{beck2024xlstmextendedlongshortterm} as well as patch-based image recognition \cite{alkin2024visionlstmxlstmgenericvision}, while also demonstrating superior sequence length extrapolation capabilities. However, the ability of xLSTMs to learn general-purpose audio representations in a self-supervised setting is yet to be thoroughly evaluated.
In this work, we propose Audio xLSTMs (AxLSTMs) for learning self-supervised general-purpose audio representations within a masked modelling framework from spectrograms patches. Pretrained on AudioSet \cite{gemmeke2017audio}, AxLSTMs consistently outperform comparable self-supervised audio spectrogram transformer (SSAST) \cite{gong2022ssast} based baselines on ten varied downstream tasks, even matching the performance of more recent and involved masked modeling approaches, such as BEATs \cite{chenbeats23}. Code and pretrained models can be found  at {\url{https://github.com/SarthakYadav/axlstm-official}}.

\begin{figure*}
    \centering
    \includegraphics[trim={2em 7.5em 1.5em 5em}, clip,width=\textwidth]{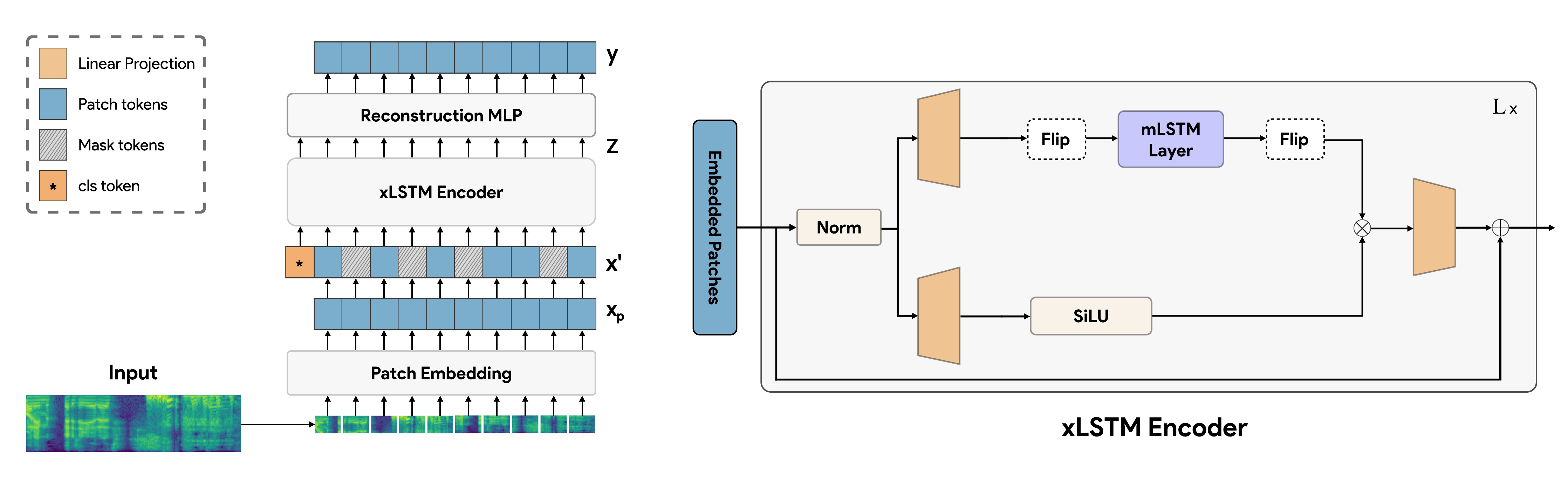}
    \caption{An overview of the proposed AxLSTM approach (left), and the constituent mLSTM blocks (right).}
    \label{fig:ssamoverview}
\end{figure*}

\section{Method}
\subsection{Prerequisites: xLSTM and the mLSTM block}
\label{ssec:prereq}

As previously discussed, xLSTM \cite{beck2024xlstmextendedlongshortterm} proposes two new building blocks, sLSTM and mLSTM. In line with Vision-LSTM \cite{alkin2024visionlstmxlstmgenericvision}, the parallelizable mLSTM module is the fundamental building block of our approach, and thus is the focus of our discussion. The mLSTM block addresses the issues with the original LSTM cell by utilizing a matrix memory cell $C \in \mathbb{R}^{d\times d}$. mLSTM stores a key-value vector pair $k_t,v_t \in \mathbb{R}^{d}$ at timestep $t$, and the relevant value vector is retrieved later using a query $q_{t+\tau}  \in \mathbb{R}^{d}$. Equations \ref{eq:mlstm_recurrent_begin}-\ref{eq:mlstm_recurrent_end} describe the forward pass for mLSTM module, where $C_t$ is the matrix memory cell, $n_t$ and $h_t$ represent the normalizer state and the hidden state, and $i_t, f_t$ and $o_t$ represent the input, forget and the output gates, respectively. Weights $W_q, W_k$ and $W_v$ are learnable projection matrices for vectors query $q$, key $k$ and value $v$, respectively. Since there is no memory mixing in mLSTMs, multiple memory cells and multiple heads are equivalent, and the forward pass can be parallelized. Further, mLSTM uses exponential gating to mitigate the inability of LSTMs to revise storage decisions when a more similar input is found, which impacts retrieval performance. 

\begin{align}
\label{eq:mlstm_recurrent_begin}
{C_t} \ &= \  {f_t} \ {C_{t-1}} \ + \ 
  {i_t} \ {v_t \ k_t^\top} ,\\
{n_t} \ &= \  {f_t} \ {n_{t-1}} \ + \ 
  {i_t} \ {k_t} ,\\
h_t  \ &= \ {o_t} \ \odot \ {C_t} {q_t} \ / \ 
\max \left\{ |{n_t^\top} {q_t}|, 1 \right\}  ,\\
q_t \ &= \ W_q \ x_t \ + \ b_q  & & \\
k_t \ &= \ \frac{1}{\sqrt{d}} W_k \ x_t \ + \ b_k  & &\\
v_t \ &= \ W_v \ x_t \ + \ b_v  & & \\
{i_t} \ &= \ \exp{\left(\ w^\top_{i} \ x_t \ + \  b_{i}  \right)} \ ,\\
{f_t} \ &= \ \exp{\left(\ w^\top_{f} \ x_t  \ + \ b_{f} \right)} ,\\
o_t \ &= \ \sigma \left( \ W_{o} \ x_t \ + \
b_{o} \right) \,
\label{eq:mlstm_recurrent_end}
\end{align}

\subsection{AxLSTM: Modelling masked patches with xLSTMs}
\label{approach}

\textbf{Creating and masking patches:} Given an input spectrogram $\mathbf{x} \in \mathbb{R}^{\mathtt{T} \times \mathtt{F}}$, the axes denoting time and frequency, respectively, we compute non-overlapping patches of shape $t \times f$, yielding $\mathbf{x}_p \in \mathbb{R}^{N \times (t \cdot f)}$ patches. Similar to \cite{dosovitskiy2021an, huang2022masked}, fixed sinusoidal 2d positional embeddings are added to the patches after projecting them to a $\mathbb{R}^{N \times d_m}$ dimensional space. To remain in line with previous work \cite{dosovitskiy2021an, gong2022ssast} and to facilitate fine-tuning of the obtained models, a representative class token is then added to the beginning of the sequence. We then proceed to randomly mask 50\% of the input patches using an unstructured masking strategy and replace these masked patches with a learnable \textit{mask} token. Thus, input to the encoder is:
\begin{equation}
    \mathbf{x'} = [\mathtt{cls}, \mathbf{x}_p^1, \mathbf{x}_p^2, \dots, \mathbf{x}_p^{N}] + E_{\mathtt{pos}}
\end{equation}
\textbf{Encoding and Reconstruction:} These partially masked patches are now fed to the encoder, yielding encoded representations $\mathbf{z} = \mathtt{enc}(\mathbf{x}'), \mathbf{z} \in \mathbb{R}^{(N+1) \times d_m}$. The encoder is a stack of mLSTM blocks as shown in Fig. \ref{fig:ssamoverview}, where each block expands the $d_m$ dimensional input by an expansion factor $E_f$ (usually $2$ \cite{beck2024xlstmextendedlongshortterm,alkin2024visionlstmxlstmgenericvision})) before projecting it back to $d_m$. The resulting blocks have fewer parameters than a comparable transformer block. To facilitate bidirectional modeling of the input, a \textit{flip} operation that reverses the order of the input sequence is enabled in even numbered mLSTM blocks, similar to \cite{alkin2024visionlstmxlstmgenericvision}. After encoding, a single hidden layer MLP is used to reconstruct patches from encoded representation $\mathbf{z}$:
\begin{align}
    \mathbf{y}' &= \mathtt{Linear}_{(t \cdot f)}(\sigma(\mathtt{Linear}_{d_m}(\mathbf{z}))),
\end{align}
where $\mathtt{Linear}_d$ is a parameterized linear projection to dimensions $d$, and $\sigma$ denotes the GELU non-linear activation function. Finally, the mean square error (MSE) between the original input spectrogram and the reconstructions, computed only for the masked patches, is used for pretraining.
Further details can be found in Section~\ref{ssec:impdetails}.
\begin{table}[h]
  \caption{Overview of downstream tasks}
  \setlength\tabcolsep{0.3pt}
  \label{tab:tasks}
  \centering
  \scriptsize
  \begin{tabular}{lcccc}
    \toprule
    ID & Name     & Description   & \#Classes & \#Hours  \\
    \midrule
    BO & Beijing Opera  & percussion instrument classification & 4 & 0.3 \\
    CD & Crema-D    & emotion recognition & 6 & 10 \\
    E50 & ESC-50  & environmental sound classification & 50 & 2.77 \\
    LC & LibriCount & counting speakers (classification) & 10 & 8  \\
    Mri-S & Mridangam Stroke & classifying Mridangam \textit{strokes} & 10 & 1.57  \\
    Mri-T & Mridangam Tonic  & classifying Mridangam \textit{tonics} & 6 & 1.57  \\
    NS-5h & NSynth Pitch 5h  & pitch classification & 88 & 5.5  \\
    SC-5h & SpeechCommands 5h  & keyword spotting & 12 & 6.5  \\
    F50K & FSD50K & multilabel audio tagging & 200 & 100  \\
    VL & VoxLingua107 Top10 & spoken language identification & 10 & 5  \\
    \bottomrule
  \end{tabular}
\end{table}
\newline\textbf{SSAST as a Baseline:} SSAST \cite{gong2022ssast} is the most directly comparable transformer-based approach to AxLSTMs. As the first patch-based self-supervised learning framework for audio spectrogram transformers (ASTs), it is a well-established method in audio representation learning. While newer transformer-based approaches such as masked autoencoders \cite{niizumi2022masked,huang2022masked,yadav2024masked} and self-distilled tokenizers in BEATs \cite{chenbeats23} introduce additional architectural modifications, SSAST offers a modular framework for evaluating the modeling capabilities of xLSTM v/s the transformer without additional confounding factors. Unlike the original SSAST, which used a combined token prediction and reconstruction based pretraining objective, we simplify the approach by using a reconstruction-only objective.
\begin{table*}[t]
    \caption{Comparing AxLSTMs with popular self-supervised audio representations. We used pretrained models from cited papers to extract fixed feature vectors and conducted our own downstream experiments. LS, AS, VP, LL stand for LibriSpeech, AudioSet, VoxPopuli and LibriLight datasets, respectively. Original SSAST \cite{gong2022ssast} was trained on AS+LS, whereas we pretrained the directly comparable underlined SSAST baselines (SSAST-T,S,B). *includes decoder parameters}
    \setlength\tabcolsep{2.25pt}
    \small
    \centering
    \begin{tabular}{lcccccccccccccc}
    \toprule
    \multicolumn{3}{l}{} &\multicolumn{4}{c}{Music \& Pitch}  &\multicolumn{4}{c}{Speech-based tasks} &\multicolumn{2}{c}{Audio}\\
    \cmidrule(lr){4-7} \cmidrule(lr){8-11} \cmidrule(lr){12-13}
    \multicolumn{1}{l}{Model} & \multicolumn{1}{l}{Data} & \#M Params & BO & Mri-S & Mri-T & NS-5h & CD & LC & SC-5h & VL & E50 & F50K & $s(m)$ \\
    \midrule
    \multicolumn{3}{l}{\textbf{Supervised Baselines}} & \multicolumn{5}{l}{}\\
    \rowcolor{Gray}
    HEAR-Naive \cite{turian_hear_2022} & - & - & 52.6\ci{2.4} & 38.0\ci{1.3} & 36.4\ci{1.9} & 18.6\ci{4.4} & 30.9\ci{0.8} & 33.5\ci{1.1} & 8.5\ci{0.4} & 11.2\ci{0.5} & 5.8\ci{0.2} & 7.1\ci{0.2} & 5.1\ci{0.7}\\
    PaSST-Base \cite{koutini22_interspeech} & AS & $86$ & 94.9\ci{0.5} & 96.5\ci{0.1} & 87.6\ci{0.6} & 23.3\ci{0.9} & 61.0\ci{0.3} & 60.1\ci{0.2} & 66.6\ci{1.4} & 25.5\ci{0.8} & \textbf{94.8\ci{0.3}} & \textbf{64.2\ci{0.1}} & 74.4\ci{0.4}\\

    \midrule
    \multicolumn{3}{l}{\textbf{SSL}} & \multicolumn{5}{l}{}\\
    \rowcolor{Gray}
    W2V2-large \cite{baevski2020wav2vec} & VP & $315.4$ & 93.1\ci{0.7} & 93.9\ci{0.1} & 77.4\ci{0.2} & 42.0\ci{1.0} & 66.9\ci{0.4} & 62.4\ci{0.3} & 87.6\ci{0.5} & 53.6\ci{1.0} & 60.1\ci{0.5} & 34.2\ci{0.1} & 74.9\ci{0.4}\\
    WavLM-large \cite{chen2022wavlm} & Mix & $315.4$ & \textbf{96.4\ci{0.5}} & 96.8\ci{0.1} & 89.5\ci{0.1} & 53.7\ci{0.5} & 57.2\ci{0.2} & 61.1\ci{0.3} & 46.2\ci{0.8} & 23.7\ci{0.9} & 47.9\ci{0.4} & 29.0\ci{0.1} & 64.8\ci{0.2}\\
    \rowcolor{Gray}
    HuBERT-large \cite{hsu2021hubert} & LL & $315.4$ & 94.1\ci{0.7} & 95.3\ci{0.1} & 83.5\ci{0.3} & 19.3\ci{0.8} & 70.7\ci{0.1} & 59.9\ci{0.2} & 83.2\ci{0.7} & \textbf{66.1\ci{0.9}} & 60.3\ci{0.4} & 31.5\ci{0.1} & 74.3\ci{0.3}\\
    BEATs-iter3 \cite{chenbeats23} & AS & $90.0$ & 94.0\ci{0.8} & 94.7\ci{0.1} & 95.8\ci{0.1} & 69.4\ci{0.8} & 67.3\ci{0.2} & 68.0\ci{0.2} & 85.2\ci{0.3} & 38.5\ci{1.0} & 83.7\ci{0.3} & 53.6\ci{0.2} & 86.7\ci{0.3}\\
    \rowcolor{Gray}
    AudioMAE \cite{huang2022masked} & AS & $86.0$ & 93.7\ci{0.6} & 89.2\ci{0.2} & 86.6\ci{0.2} & 64.5\ci{0.8} & 68.2\ci{0.2} & 42.2\ci{0.2} & 28.6\ci{1.5} & 29.7\ci{1.0} & 60.6\ci{0.4} & 37.9\ci{0.1} & 63.6\ci{0.3}\\
    MSM-MAE-208 \cite{niizumi2022masked} & AS & ${92.7}^{*}$ & 95.7\ci{0.7} & 97.3\ci{0.1} & \textbf{97.9\ci{0.1}} & 69.1\ci{0.5} & 68.7\ci{0.2} & 63.8\ci{0.5} & 85.7\ci{0.3} & 40.3\ci{0.6} & 78.4\ci{0.6} & 49.5\ci{0.1} & 86.0\ci{0.2}\\
    \rowcolor{Gray}
    MWMAE-Tiny \cite{yadav2024masked} & AS & ${12.6}^{*}$ & 93.3\ci{1.0} & 97.1\ci{0.1} & 97.6\ci{0.1} & 68.1\ci{0.4} & 64.4\ci{0.2} & 65.5\ci{0.3} & 77.0\ci{0.6} & 28.6\ci{1.1} & 71.9\ci{0.5} & 43.4\ci{0.1} & 80.0\ci{0.3}\\
    MWMAE-Base \cite{yadav2024masked} & AS & ${92.5}^{*}$ & 96.0\ci{0.5} & 97.4\ci{0.1} & \textbf{97.9\ci{0.1}} & 69.3\ci{0.6} & \textbf{73.1\ci{0.3}} & \textbf{68.8\ci{0.2}} & \textbf{90.9\ci{0.2}} & 44.2\ci{0.9} & 81.2\ci{0.4} & 51.2\ci{0.2} & \textbf{90.3\ci{0.2}}\\
    \rowcolor{Gray}
    SSAM-Tiny \cite{yadav2024audiomambaselectivestate} & AS & $4.8$ & 93.7\ci{0.8} & 97.1\ci{0.1} & 94.9\ci{0.1} & 62.0\ci{0.7} & 61.8\ci{0.3} & 59.2\ci{0.4} & 74.8\ci{0.4} & 27.8\ci{1.0} & 70.6\ci{0.2} & 41.3\ci{0.2} & 75.6\ci{0.2}\\
    SSAM-Small \cite{yadav2024audiomambaselectivestate} & AS & $17.9$ & 94.0\ci{0.7} & 97.5\ci{0.1} & 96.7\ci{0.1} & 66.3\ci{0.8} & 67.5\ci{0.2} & 60.5\ci{0.3} & 83.7\ci{0.3} & 39.6\ci{0.7} & 78.7\ci{0.6} & 48.5\ci{0.1} & 83.5\ci{0.3}\\
    \rowcolor{Gray}
    SSAM-Base \cite{yadav2024audiomambaselectivestate} & AS & $69.3$ & 93.2\ci{1.1} & \textbf{97.7\ci{0.1}} & 96.9\ci{0.1} & 70.5\ci{0.5} & 70.3\ci{0.2} & 63.5\ci{0.2} & 87.9\ci{0.3} & 50.4\ci{0.7} & 81.0\ci{0.3} & \underline{52.2\ci{0.1}} & 88.7\ci{0.3}\\
    \midrule
    \multicolumn{3}{l}{\textbf{SSAST Based}} & \multicolumn{5}{l}{}\\
    \rowcolor{Gray}
    SSAST \cite{gong2022ssast} & Mix & $89.0$ & 93.4\ci{0.9} & 96.7\ci{0.1} & 96.3\ci{0.1} & 66.8\ci{0.7} & 56.5\ci{0.2} & 60.7\ci{0.3} & 53.5\ci{1.3} & 28.5\ci{0.9} & 68.4\ci{0.4} & 38.2\ci{0.1} & 72.5\ci{0.2}\\
    
    \underline{SSAST-T} & AS & $5.4$ & 90.4\ci{0.7} & 95.7\ci{0.1} & 94.3\ci{0.1} & 61.2\ci{0.5} & 46.9\ci{0.2} & 42.7\ci{0.2} & 50.6\ci{1.6} & 13.8\ci{1.0} & 42.4\ci{0.6} & 24.6\ci{0.1} & 55.6\ci{0.2}\\
    \rowcolor{Gray}
    \underline{SSAST-S} & AS & $21.5$ & 93.2\ci{0.5} & 96.2\ci{0.1} & 95.0\ci{0.1} & 63.8\ci{0.4} & 51.6\ci{0.2} & 50.0\ci{0.3} & 58.3\ci{1.2} & 15.6\ci{0.7} & 50.1\ci{0.6} & 31.6\ci{0.1} & 63.0\ci{0.2}\\
    
    \underline{SSAST-B} & AS & $85.7$ & 93.1\ci{0.7} & 96.6\ci{0.1} & 96.2\ci{0.2} & 64.6\ci{0.8} & 56.0\ci{0.4} & 52.9\ci{0.3} & 66.1\ci{1.0} & 19.2\ci{0.9} & 59.6\ci{0.7} & 37.5\ci{0.1} & 68.7\ci{0.3}\\
    \midrule
    \multicolumn{3}{l}{\textbf{Proposed}} & \multicolumn{5}{l}{}\\
    \rowcolor{Gray}
    AxLSTM-T & AS & $4.3$ & 93.9\ci{0.7} & 96.8\ci{0.1} & 94.9\ci{0.1} & 62.5\ci{0.6} & 57.5\ci{0.2} & 55.1\ci{0.5} & 69.0\ci{1.6} & 24.1\ci{0.6} & 61.3\ci{0.6} & 37.5\ci{0.2} & 70.7\ci{0.2}\\
    AxLSTM-S & AS & $16.7$ & 92.9\ci{1.0} & 97.4\ci{0.1} & 96.6\ci{0.1} & 66.6\ci{0.4} & 65.0\ci{0.2} & 60.3\ci{0.3} & 80.5\ci{0.4} & 36.5\ci{0.7} & 75.5\ci{0.4} & 46.5\ci{0.1} & 81.1\ci{0.3}\\
    \rowcolor{Gray}
    AxLSTM-B & AS & $65.6$ & 93.6\ci{0.9} & 97.5\ci{0.1} & 97.5\ci{0.1} & 71.4\ci{0.8} & 68.7\ci{0.2} & 63.2\ci{0.3} & 85.1\ci{0.2} & 43.5\ci{0.6} & 79.2\ci{0.6} & \underline{51.0\ci{0.1}} & 86.6\ci{0.2}\\
    \bottomrule
    \end{tabular}
    \label{tab:overall}
\end{table*}
\section{Experiments}
\label{sec:exp}

\subsection{Datasets}
\textbf{Pretraining:} For pretraining, we use the AudioSet dataset (AS) \cite{gemmeke2017audio} which has roughly 2 million 10-second long weakly labeled YouTube clips and over 5000 hours of audio data spanning 527 classes.
\newline\textbf{Downstream Evaluation:} Following recent work \cite{yadav2024masked, yadav2024audiomambaselectivestate}, we use the following subset of tasks proposed as part of the HEAR benchmark \cite{turian_hear_2022} for downstream evaluation: Beijing Opera \cite{beijingopera, turian_hear_2022}, Crema-D \cite{cao2014crema}, ESC-50 \cite{esc50}, LibriCount \cite{stoter2018classification}, Mridangam Stroke and Tonic \cite{mridangamds}, NSynth Pitch 5h \cite{turian_hear_2022, nsynth2017}, Speech Commands 5h \cite{turian_hear_2022, warden2018speech}, FSD50K \cite{fonseca2021fsd50k} and VoxLingua107 \cite{kim2018vocal}. Table~\ref{tab:tasks} provides a brief overview of these tasks.

\subsection{Implementation details}
\label{ssec:impdetails}

\textbf{Spectrogram features:} We extract log-scaled mel spectrogram features with a window size of $25$ ms, a hop size of $10$ ms and $F=80$ mel-spaced frequency bins. A sampling rate of $16000$ Hz is used for all audio clips.\newline
\textbf{Pretraining:} All the proposed models are pretrained on randomly cropped 2-second audio clips, resulting in an input spectrogram of shape $[\mathtt{200} \times \mathtt{80}]$. 
Our default configuration consists of $l=12$ number of stacked mLSTM blocks with a model feature dimension of $d_m=192$, the same as those of a ViT-Tiny \cite{dosovitskiy2021an} encoder, but we also evaluate Small ($d_m=$384, $l$=12) and Base ($d_m$=768, $l$=12) encoder configurations. 
By default, a patch embedding layer that computes $[4\times16]$ shaped non-overlapping patches (along time and frequency, respectively) is used in all our experiments. 
While \cite{beck2024xlstmextendedlongshortterm, alkin2024visionlstmxlstmgenericvision} suggest a default expansion factor of $E_f=2$, in early ablations we found that $E_f=3$ performed the best (Section \ref{ssec:ablations}). Ablation experiments also revealed that disabling the flip operation performed better than enabling flip in alternating blocks. 
We build on hyperparameter recommendations from previous work \cite{gong2022ssast,yadav2024audiomambaselectivestate}: all models are trained for 100 epochs with a batch size of 1024 and a weight decay of 0.05. AdamW optimizer with a linear warmup for 10 epochs followed by a cosine learning rate decay schedule is used. No data augmentations were used.\newline
\textbf{Downstream evaluation:} Following the HEAR protocol, we extract fixed-sized feature vectors independent of the input audio duration by taking the mean over time across 2-second audio chunks. We then train a single hidden layer MLP classifier with 1024 neurons for each task, using the official \textit{hear-eval-kit} accompanying the HEAR benchmark. All experiments are repeated with 10 different random seeds, and 95\% confidence intervals are reported. It is worth noting that due to the number of experiments conducted, we use a restricted hyperparameter grid (same as \cite{yadav2024audiomambaselectivestate,yadav2024masked}) compared to the \textit{hear-eval-kit}. \newline
\textbf{Aggregated Performance Metric:} We use the aggregated normalized score as proposed by \cite{yadav2024masked} to compare evaluated approaches across the proposed list of downstream tasks. For a model $m$, overall score $s(m) \in [0,100]$ is given as: $s(m) = \frac{1}{|T|}\sum_{t \in T} \frac{x_t(m) - min_t}{max_t - min_t} * 100$, where $x_t(m)$ denotes performance of the model $m$ on task $t$, and $min_t$ and $max_t$ represent the worst and the best performance across all models on the task, thus taking into account the relative performance amongst all evaluated representations.

\section{Results}
\label{sec:exps}

\subsection{Comparison with existing works}

Table~\ref{tab:overall} shows how the proposed AxLSTM models fare against recent audio representations. Directly comparable SSAST and AxLSTM feature representations have identical feature embedding sizes, however, feature vector sizes extracted using other referred methods can vary. While sub-optimal, it is infeasible to retrain all pretrained representations to have the same embedding sizes, and our evaluation protocol is in line with recent frameworks for evaluating self-supervised audio representations \cite{turian_hear_2022, yang21c_interspeech}.
\textit{SSAST \cite{gong2022ssast}} represents the officially released model trained on AudioSet+LibriSpeech, whereas underlined SSAST representations were pretrained by us. We can see that AxLSTM models consistently outperform their transformer-based SSAST counterparts by a considerable margin, while having over 23\% fewer parameters, with AxLSTM-B yielding over 25\% relative improvement in aggregate performance (86.6{\ci{0.2}} v/s 68.7{\ci{0.3}}). 
Despite being based on the older SSAST framework, our AxLSTM-B model matches the more recent masked modeling-based approaches such as AudioMAE \cite{huang2022masked}, BEATS-iter3 \cite{chenbeats23} and MSM-MAE-208 \cite{niizumi2022masked} in overall performance, while having over 25\% fewer parameters. The recent Mamba-based SSAM \cite{yadav2024audiomambaselectivestate} models outperform comparable AxLSTM models in overall performance, however, it’s worth noting that AxLSTM models outperform SSAM models in music and pitch perception tasks (BO, Mri-T, NSynth-5h) whereas SSAM models shine in Speech-based and Audio classification tasks, performing notably better for emotion recognition (CD) and spoken language identification (VL). Finally, the recent MWMAE-Base \cite{yadav2024masked} model with local-global self-attention outperforms every evaluated approach, including AxLSTM-B, albeit with notably more parameters. Overall, while the AxLSTM model doesn't establish a new state-of-the-art, it significantly outperforms SSAST-based transformer baselines while matching the performance of several other newer masked modeling approaches while possessing fewer parameters, thus demonstrating the viability of xLSTMs to learn self-supervised audio representations.

\subsection{Ablations}
\label{ssec:ablations}
This section covers ablations for investigating the impact of key configurable hyperparameters. Unless stated otherwise, experiments use the Tiny configuration with an expansion factor $E_f=2$ and flipping enabled in alternating mLSTM blocks. 
\newline\textbf{Expansion Factor:} This hyperparameter affects the size and modeling capacity of mLSTM blocks. Table \ref{tab:expfac} shows that $E_f=3$ offers better overall performance than the default $E_f=2$ (suggested in \cite{beck2024xlstmextendedlongshortterm, alkin2024visionlstmxlstmgenericvision}). However, $E_f=2$ still outperforms the SSAST baseline, offering 20\% better performance with over 45\% fewer parameters.
\begin{table}[t]
    \caption{Performance impact of expansion factor ($E_F$)}
    \small
    \label{tab:expfac}
    \setlength\tabcolsep{5pt}
    \centering
    \begin{tabular}{l|ccc}
        \toprule
        Model & $E_f$ & \# Params & $s(m)$\\
        \midrule
        SSAST-T & - & 5.4 M & 55.6\ci{0.2}\\
        \midrule
        AxLSTM-T & 2 & 2.9 M & 66.9\ci{0.3} \\
        AxLSTM-T & 3 & 4.3 M & \textbf{68.9\ci{0.4}} \\
        AxLSTM-T & 4 & 5.8 M & 63.4\ci{0.2}\\
        \bottomrule
    \end{tabular}
\end{table}
\newline\textbf{Patch Size:} The patch size hyperparameter governs the sequence length as well as the time-frequency resolution of the patches that are processed by the underlying sequence modeling blocks. We pretrain and compare SSAST and AxLSTM Tiny configurations with 3 patch sizes: $(4,8)$, $(4,16)$, and $(8, 16)$. From Table~\ref{tab:patchablations}, we can see that AxLSTM scales better with increasing number of patches, with smaller patch sizes continually leading to better performance, whereas SSAST-T experiences a drop in performance at patch size of $(4,8)$, suggesting that AxLSTMs scale better with sequence length in terms of modeling performance compared to standard transformers.
\begin{table}[t]
    \caption{Patch size ablations with the Tiny configuration}
    \small
    \label{tab:patchablations}
    \setlength\tabcolsep{5pt}
    \centering
    \begin{tabular}{l|ccc}
        \toprule
        Model & Patch Size & \# Patches & $s(m)$\\
        \midrule
        SSAST-T & $(8,16)$ & 125 & 47.0\ci{0.3} \\
        AxLSTM-T &  $(8,16)$ & 125 & \textbf{55.2\ci{0.3}} \\
        \midrule
        SSAST-T & $(4,16)$ & 250 & 55.6\ci{0.2} \\
        AxLSTM-T &  $(4,16)$ & 250 & \textbf{66.9\ci{0.3}} \\
        \midrule
        SSAST-T & $(4,8)$ & 500 & 53.8\ci{0.3} \\
        AxLSTM-T & $(4,8)$ & 500 & \textbf{68.2\ci{0.3}} \\
        \bottomrule
    \end{tabular}
\end{table}
\newline\textbf{Bidirectional operation and exponential gating:} 
In Table \ref{tab:various}, we evaluate the impact of bidirectional modeling (controlled through the \textit{flip} operation) and replacing exponential gating with a sigmoid gate. When done one at a time, disable both flip operation in alternating xLSTM blocks and the exponential gate improve performance compared to when both of these operations are enabled, with the former leading to a greater performance improvement. This suggests that bidirectional modeling is not as important for downstream audio classification as it is for vision \cite{alkin2024visionlstmxlstmgenericvision}, which is inline with observations made in previous work \cite{yadav2024audiomambaselectivestate}. However, disabling both exponential gating and flipping together does not further improve performance, instead it performs worse than just disabling sequence flipping in alternative blocks. These results suggest that changes at the architectural level to accommodate bidirectional modeling and exponential gating might be warranted. It is also worth noting that the best performing AxLSTM-T model with $E_f=2$ outperforms the SSAST-T baseline by 26\% but has 45\% fewer parameters.

\begin{table}[ht]
    \caption{Evaluating the impact of exponential gating and flipping on overall performance}
    \small
    \label{tab:various}
    \centering
    \begin{tabular}{l|cccc}
    \toprule
    Model & $E_f$ & Exp gate & Flip  & $s(m)$     \\
    \midrule
    AxLSTM-T & 2 & \checkmark & \checkmark  & 66.9±0.3 \\
    AxLSTM-T & 2 & \checkmark &  & 70.1±0.3 \\
    AxLSTM-T & 2 &  & \checkmark  & 68.2±0.2 \\
    \midrule
    AxLSTM-T & 3 & \checkmark & \checkmark  & 68.8±0.4 \\
    AxLSTM-T & 3 & \checkmark &  & 70.7±0.2 \\
    AxLSTM-T & 3 &  &  & 69.6±0.2\\
    \bottomrule
    \end{tabular}
\end{table}

\section{Conclusion}

This work presents AxLSTMs, an approach that evaluates the viability of the recently proposed xLSTM model for learning self-supervised audio representations from masked audio spectrograms. Extensive empirical analysis shows that AxLSTMs significantly outperform comparable SSAST baselines on ten varied downstream audio recognition tasks. AxLSTM-B, which is our best performing model, offers over 25\% better performance than SSAST-B while having up to 23\% fewer parameters, and even matches the performance of several newer masked modeling approaches such as MSM-MAE-208 and BEATs. On the other hand, AxLSTM-T models with $E_f=2$ can offer over 25\% relative improvement compared to an SSAST baseline, while having over 45\% fewer parameters, which is a great performance/complexity tradeoff. Overall, our results demonstrate the viability of xLSTMs for learning self-supervised audio representations.




\newpage
\bibliographystyle{IEEEtran}
\bibliography{mybib}

\end{document}